\documentclass[aps,pre,floats,superscriptaddress,twocolumn]{revtex4}

\usepackage{graphicx}
\usepackage{graphics}
\graphicspath{{Figures/}} 
\usepackage{amsmath}
\usepackage{amsfonts}
\usepackage{amssymb}
\usepackage{epstopdf}
\usepackage{makeidx}
\usepackage{epsfig}
\usepackage{bm}
\usepackage[colorlinks=true,citecolor=blue,linkcolor=blue,linktocpage=true,pagebackref=false]{hyperref}
\usepackage{color,soul} 
\usepackage{float}
\usepackage[english]{babel}

\newcommand{\ie}{\textit{i.e.}\,}
\newcommand{\be}{\begin{equation}}
\newcommand{\ee}{  \end{equation}}
\newcommand{\ba}{\begin{eqnarray}}
\newcommand{\ea}{  \end{eqnarray}}

\begin{document}
	
	\title{
Polarization 
inhibits the phase transition of Axelrod's model}	

	\author{Carlos Gracia-L\'azaro}
	\affiliation{Instituto de Biocomputaci\'on y F\'{\i}sica de Sistemas Complejos (BIFI), Universidad de Zaragoza, 50018 Zaragoza, Spain}
	\author{Edgardo Brigatti}
	\affiliation{Instituto de F\'{\i}sica,
		Universidade Federal do Rio de Janeiro, 22452-970 Rio de Janeiro, Brazil}
	\author{Alexis R. Hern\'andez}
	\affiliation{Instituto de F\'{\i}sica,
		Universidade Federal do Rio de Janeiro, 22452-970 Rio de Janeiro, Brazil}
	\author{Yamir Moreno}
	\affiliation{Instituto de Biocomputaci\'on y F\'{\i}sica de Sistemas Complejos (BIFI), Universidad de Zaragoza, 50018 Zaragoza, Spain}
	\affiliation{Departamento de F\'{\i}sica Te\'orica. Universidad de Zaragoza, Zaragoza E-50009, Spain}
	\affiliation{ISI Foundation, Turin, Italy}
	
	\date{\today}
	
	\begin{abstract}
We study the effect of polarization in Axelrod's model of cultural dissemination.
This is done through the introduction of a cultural feature that takes only
two values, while the other features can present a larger number of possible traits.
Our numerical results and mean-field approximations show that polarization reduces
the characteristic phase transition of the original model to a finite-size effect,
since at the thermodynamic limit only the ordered phase is present.
Furthermore, for finite system sizes, the stationary state depends on the percolation
threshold of the network where the model is implemented:
a polarized phase is obtained for percolation thresholds below 1/2,
a fragmented multicultural one otherwise.	
	

	\end{abstract}

	\maketitle

\section{Introduction}

Agent-based models (ABM) \cite{macy2002factors,tesfatsion2006handbook} provide a fruitful theoretical framework to study the fundamental mechanisms underlying the dynamics of social systems \cite{castellano2009statistical}. In this line, the Axelrod's Model for cultural dissemination \cite{axelrod1997dissemination}, introduced in 1977 by Robert Axelrod, has become a paradigm for the study of social imitation. The model relays on the idea of homophily, \ie, agents interact more likely with similar neighbors, and therefore, similar neighbors tend to become even more alike. To implement this idea in the model, the probability for an agent to imitate a neighbor's uncommon cultural trait is proportional to the number of other traits that both already share. Axelrod found that, while for low values of the initial cultural diversity the dynamics drives the system towards a monocultural state, for larger initial diversity the system freezes in a multicultural state.
The  Axelrod's Model has been studied under different approaches and variations, including 
complex networks \cite{klemm2003nonequilibrium,guerra2010dynamical}, clustering \cite{lanchier2012axelrod},
social pressure \cite{kuperman2006cultural}, noise \cite{klemm2003global,klemm2005globalization}, external fields \cite{gonzalez2005nonequilibrium,gonzalez2007information}, dynamic features \cite{hernandez2018robustness}, mobility and segregation \cite{gracia2009residential,pfau2013co}, tolerance \cite{gracia2011selective}, confidence 
thresholds \cite{de2009effects}, and dynamic networks \cite{gracia2011coevolutionary}. 

In this work, we are interested in adding to the Axelrod's model an element that can account for polarization. Cultural, ideological, and political polarizations are phenomena that recently have attracted the attention of the scientific community, politics, and society as a whole \cite{conover2011political,prior2013media,baldassarri2007dynamics,dalton2006social}. A source of this renewed interest stems from the observation that, paradoxically, polarization may constitute a side effect of globalization. Although according to the homogenization thesis, globalization should standardize a global pattern \cite{levin2001globalizing,hallin2004americanization,gordon2009globalization}, cultural alternatives and resistance to Western norms suggest that culture universalization may lead to polarization \cite{ksiazek2008cultural,flache2011small}. These two theses constitute the convergence-divergence open debate \cite{holton2000globalization}. To this end, by introducing a cultural feature with only two possible values (\textit{e.g.}, left-right, east-west), 
we propose a modification of the Axelrod's model that allows exploring the consequences of polarizing issues on cultural diversity.

The introduction of a polarizing issue in the Axelrod's model seems to have an important
impact on the existence of its phase transition.
The order-disorder transition described by Castellano {\it et al.} \cite{castellano2000nonequilibrium} is a 
genuine phase transition which occurs in the limit of infinite system size \cite{vilone2002ordering,vazquez2007non}.
Anyway, it has already been questioned in relation to its robustness. 
In some circumstances, it has been proven that exogenous perturbations can drive the system to a monocultural state \cite{klemm2005globalization}. 
Another interesting example is the effect that cultural drift,  
modeled as a noise which randomly changes one agent's cultural trait, 
has on the model. In this case, even if a finite system shows a well defined noise value which separates the transition between order and disorder, an infinite system always results in a multicultural state \cite{klemm2005globalization,klemm2003global}.
The disappearance of the transition is also recorded when 
the original Axelrod's model is embedded on a 
Barab\'asi-Albert network \cite{klemm2003nonequilibrium}. In that case, the location of the finite-size transition point scales as a
power of the system size with a positive exponent, and, therefore, in the thermodynamic limit, the transition disappears because the ordered monocultural state always establishes in the system.
Keeping in mind this interesting aspects of the Axelrod's model, the principal aim of this work is to analyze the effects of the introduction of cultural polarization on its phase transition and to characterize the impact of this new element on the dynamics of cultural dissemination.

\section{Model}

The original Axelrod's model of cultural dissemination considers $N$ agents interconnected by a network whose links represent the social interactions. For an agent $i$, the culture is represented by a set of $F$ variables $\{\sigma_f(i)\}$ ($f=1,...,F$), the {\em cultural features}, that can assume $q$ values, $\sigma_f=0,1,...q-1$, the {\em traits} of the feature. The number of possible traits, $q$, represents the initial cultural diversity, which is obtained by means of an equiprobable random initialization of each agent's features.
At each time step, an agent $i$ is chosen at random and allowed to imitate an uncommon feature's trait of a randomly chosen neighbor $j$ with a probability given by 
their cultural overlap $\omega_{ij}$, which is defined as the fraction of common cultural features: 
\begin{equation}
\omega_{ij} =
\frac{1}{F}\sum_{f=1}^{F}\delta_{\sigma_f(i),\sigma_f(j)}\;\;,
\label{overlap}
\end{equation}
here $\delta_{x,y}$ stands for the Kronecker's delta, defined as $\delta_{xy}=1$
if $x=y$ and $\delta_{xy}=0$ otherwise. 

In this work, we will consider the situation where one of the $F$ features, $f=1$, is limited to take only two values ($\sigma_1(i)\in\{0,1\},\;\forall i\in\{1,2,\ldots,N\}$). The rest of the features, $f=2,3,\ldots,F$, can take $q$ possible values, as in the original model. In this way, we will address how a dichotomy can impact the process of cultural dissemination described by the Axelrod's model.

\section{Results}

The principal aim of our study is the characterization  of the phase transition exhibited by the model.
To this end, we simulate the model on a regular two-dimensional  square lattice of size $L$ (\ie, $N=L^2$ nodes, each node being occupied by a cultural agent) with periodic boundary conditions. We consider von Neumann neighborhoods so that each agent has $k=4$ neighbors, and interactions take place only between two neighbors. The number of features is fixed to $F=10$ and the parameters $q$ and $N$ are varied.

The phase transition is determined by the passage from an
absorbing monocultural state composed by a single cultural cluster
where,  for small $q$ values, everybody shares the same culture 
to a frozen disordered and fragmented state. 
In fact, for large values of $q$ order is not attained and different cultures, distributed
among the sites, characterize the system.

\begin{figure}[h!]
	\includegraphics[width=0.485\textwidth]{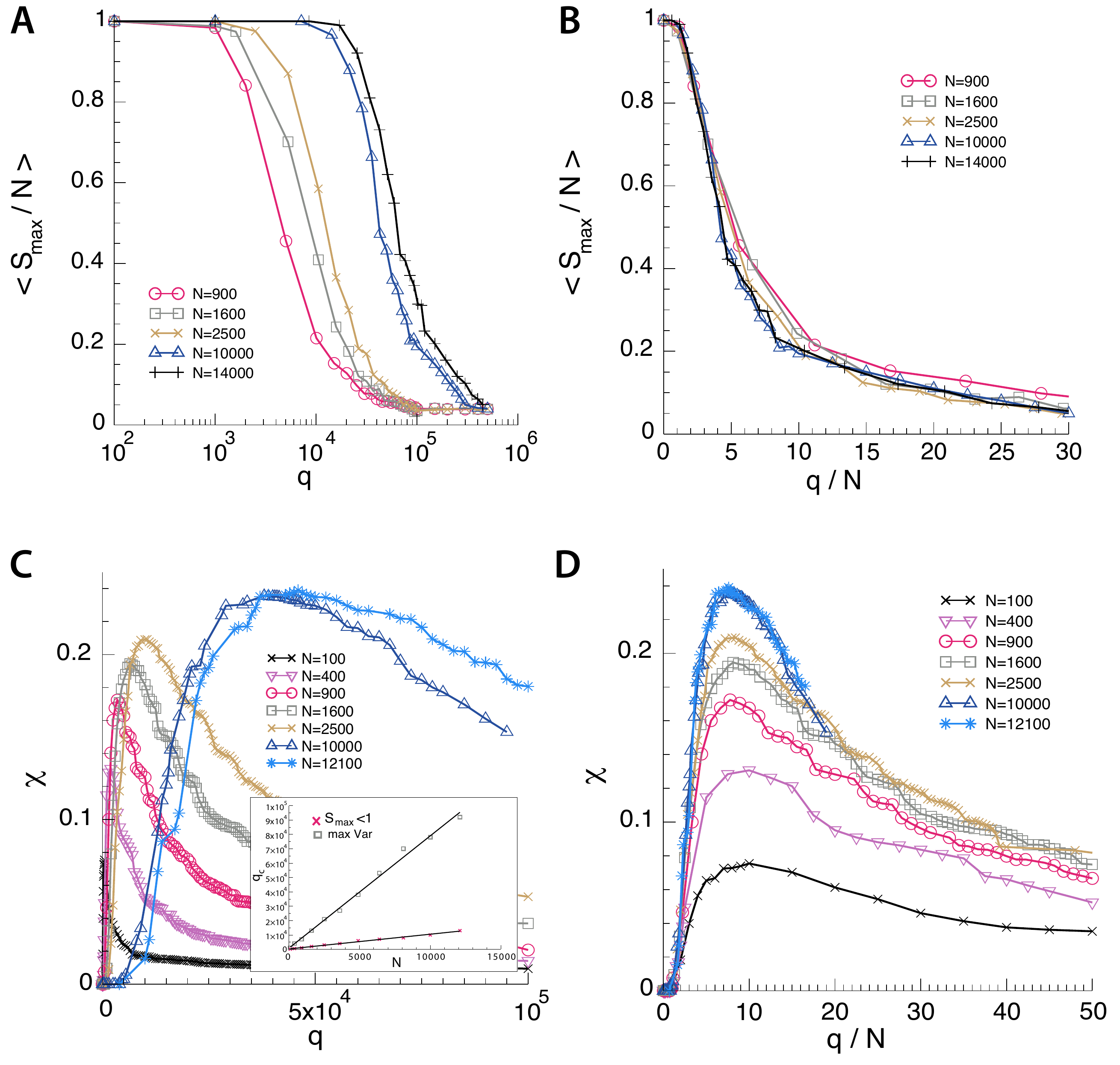}
	\caption{\textbf{A}: Mean of the normalized largest cultural cluster size $\langle S_{max}/N\rangle$ as a function of the initial cultural diversity $q$ for different system sizes $N$. \textbf{B}: The same plot after rescaling the x-axis for $q/N$. \textbf{C}: Variance $\chi$ of the order parameter $\langle S_{max}/N\rangle$ versus $q$ for different system sizes. In the inset: Value of $q_c$ for the maximum variance $\chi$ (blue squares) and for the jump location in the normalized size of the largest cluster (red Xs) as a function of the system size $N$, showing the tendency of the finite-size transition points $q_c(N)$ in the thermodynamic limit: $q_c=\lim_{N \to \infty} q_c(N)=\infty$.  \textbf{D}: $\chi$ versus $q/N$ for different system sizes $N$. It is shown that the maximum of the fluctuations in the order parameter takes place for the same value of $q/N$ regardless of the system size. 
	All these results correspond to a lattice (k=4) and F=10. Each point is averaged over 1000 simulations. 
	}
		\label{fig:transition}
\end{figure}

Because of this phenomenology, the relative size of the largest cultural cluster present in the system is an excellent parameter for characterizing the transition \cite{castellano2000nonequilibrium,brigatti2016finite,crokidakis2015discontinuous}.
This parameter is defined as the size of the largest cluster, made up by nodes
sharing all the traits, normalized by the system size: $S_{max}/N$. 
This parameter is estimated 
averaging over different simulations: $\langle S_{max}/N\rangle$.


Panel \textbf{A} of Fig. \ref{fig:transition} shows the order parameter $\langle S_{max}/N\rangle$ versus the initial cultural diversity $q$ for different system sizes $N$. The inspection of the behavior of the order parameter suggests the existence of a transition. In fact, a sharp transition, characterized by a drop of $\langle S_{max}/N\rangle$ for a critical value of $q$, is observed.  
We can identify this transition point looking at  the fluctuations of the size of the largest cluster:

\begin{eqnarray}
\chi=\langle S_{max}^{2}\rangle - \langle S_{max} \rangle^{2}\;\;.
\nonumber
\end{eqnarray}

Panel \textbf{C} of Fig. \ref{fig:transition} shows the values of $\chi$ as a function of $q$ for different system sizes $N$. This quantity  displays its maximum at a value of $q$ that can be considered as the finite-size transition point $q_c(N)$. 
From this analysis we can estimate the critical point by looking at the convergence of the finite-size transition points $q_c(N)$, as estimated by the localization of the maxima of the fluctuations. The inset of panel \textbf{C} shows that the finite-size transition point values grow proportionally to the 
system size as:   $q_c(N) \propto N$.
The same conclusion is obtained measuring the transition points from
the location of the observed discontinuity in the 
normalized size of the largest cluster. This point corresponds to the first value of $q$ for which 
$\langle S_{max}/N\rangle$ is less than 1. 
Using these results, we can obtain a data collapse by introducing the rescaled parameter $q/N$ (see panels \textbf{B} and \textbf{D} of Fig. \ref{fig:transition}). 

This scaling indicates that, in the thermodynamic limit, the transition point ($q_c=\lim_{N \to \infty} q_c(N)$) goes to infinite.  
Therefore, the system does not display a 
genuine phase transition, which is rigorously defined at the thermodynamic limit, where the number of constituents 
tends to infinity.
The introduction of the binary feature destroys the well known phase transition
of the Axelrod's model:  in the thermodynamic limit the transition disappears and the ordered monocultural state is the sole phase displayed by the system.



\begin{figure}[h!]
	\includegraphics[width=0.485\textwidth]{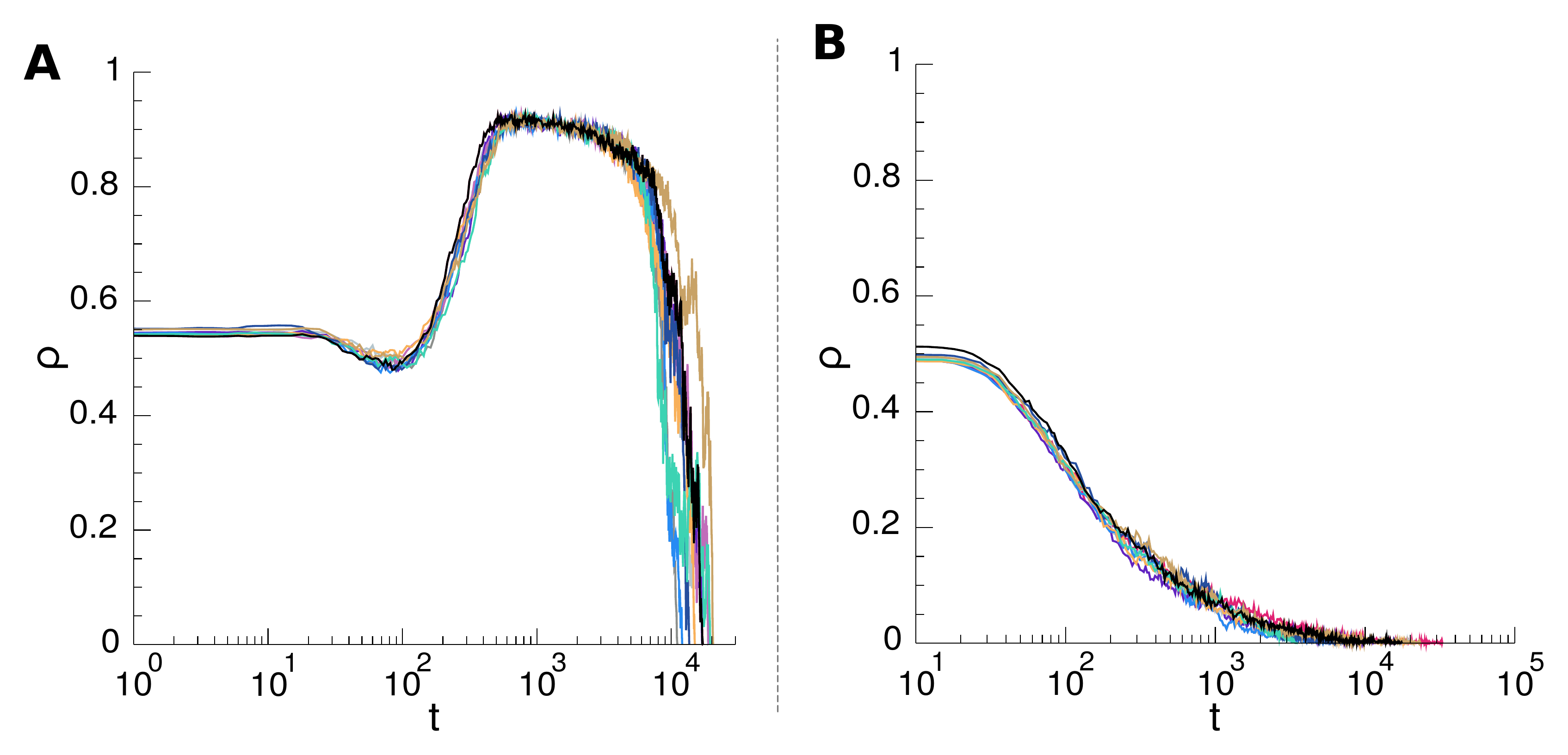}
		\caption{ Evolution of the density $\rho$ of active links in a regular lattice ($N = 1600$, $k=4$) for $q = 100$ (\textbf{A}), $q = 5 \times 10^5$ (\textbf{B}). 
Each color corresponds to a realization among ten randomly selected.
	}
	\label{fig:active}
\end{figure}

A clear understanding of this result can come from  the  exploration of the dynamics of the density $\rho$ of active links. 
An active link is a bond that connects two agents ($i,j$) with at least one different feature and at least another one equal (\ie, $0<\omega_{ij}<1$). 
The nodes at both ends of these links are the only ones that can change their states and produce some dynamics.  For this reason, $\rho=0$ implies a frozen configuration.
Figure \ref{fig:active} displays the time evolution of this quantity. 
Panel \textbf{A} of Figure \ref{fig:active} shows the result of ten different simulations for 
$q<q_c(N)$, which generate ordered final configurations. 
In this situation, $\rho$ starts near 0.5, experiences a decrease but then rises again towards a peak
before finally decaying to zero. 
The second part of the dynamics, after the maximum is reached, is a coarsening process which follows different erratic 
paths and abruptly reaches zero. Note that  the final state corresponding to one region of ordered features is reached 
as the result of a fluctuation in a finite system. A finite size effect
produces a final ordered region of size comparable to the whole system.

When $q>q_c(N)$ (Panel \textbf{B} of Figure \ref{fig:active}) the simulations do not converge to an ordered state and $\rho$ 
follows quite regular paths towards zero. 
This smooth coarsening process gives rise to regions clearly smaller than the system size (of the order of $N/10$) and produces regular similar trajectories of $\rho(t)$.
These results are totally analogous to those of the original Axelrod's model \cite{castellano2000nonequilibrium}.

\begin{figure}[h!]
	\includegraphics[width=0.3\textwidth]{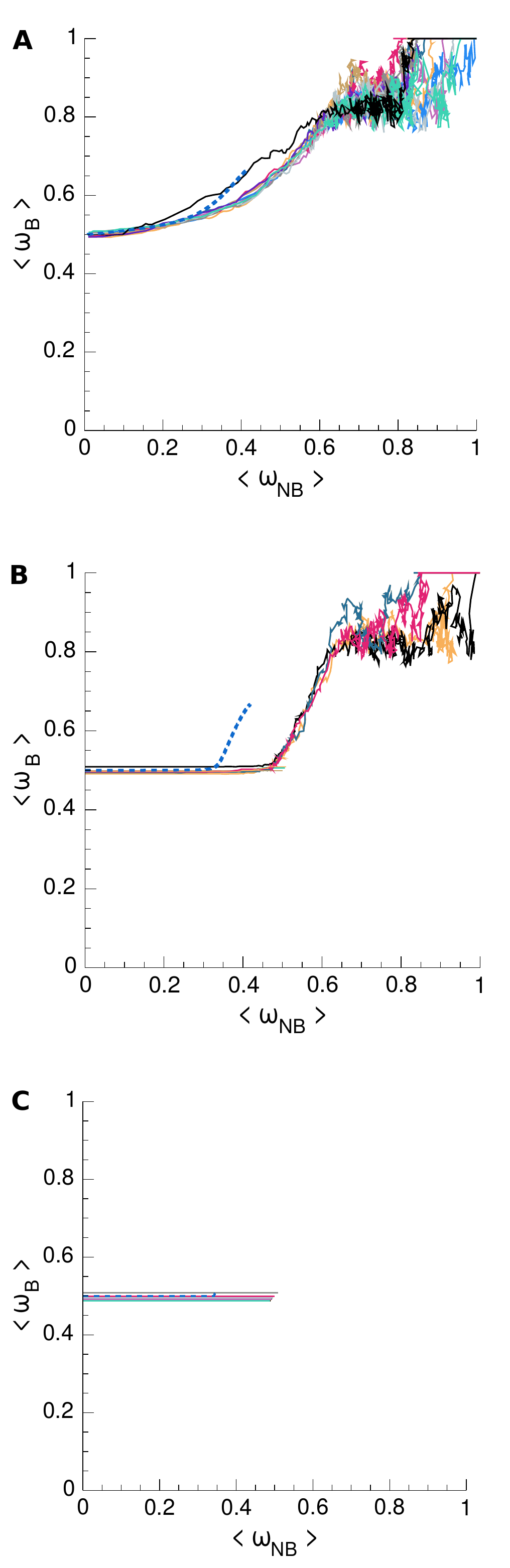}
	\caption{Evolution of the mean overlap for the binary feature versus the overlap for the non-binary features. Solid lines correspond to the numerical results for a regular lattice ($k=4$, $N = 1600$) and $q = 10^2$ (\textbf{A}), $10^4$ (\textbf{B}), $10^{12}$ (\textbf{C}). Different colors correspond to 10 different characteristic realizations. Dashed blue lines show the results corresponding to the mean-field approximation given in Eq. (\ref{eq:MF}). 
	}
	\label{fig:Overlap_vs_binary_Lattice_q}
\end{figure}

In order to better understand the impact of the binary feature on the evolution of the system, we have studied the dynamics of 
the mean overlap $\langle \omega(t) \rangle$. Panels \textbf{A}-\textbf{C} of Figure \ref{fig:Overlap_vs_binary_Lattice_q} display the trajectories of the mean overlap of the binary feature ($\langle \omega_{B}(t) \rangle$) versus the mean overlap of the non-binary features ($\langle \omega_{NB}(t) \rangle$). Solid lines correspond to the numerical results, different colors corresponding to different single simulations.  
We can distinguish three different stages. 
During the first stage, the binary feature remains almost unaltered, in contrast to the non-binary overlap which increases. 
This suggests that the cultural uniformization is realized predominantly within clusters of agents that share the binary trait.
For $q>q_c(N)$ (Panel \textbf{C}), only this first stage takes place, and the system remains frozen  in a disordered state with a characteristic number of different cultural clusters. 
For $q\lesssim q_c(N)$ (Panels \textbf{A}-\textbf{B}), after a particular overlap value of the non-binary feature is reached,
the two overlaps start to increase together. In this regime, the cultural exchange might take place also between agents marked by distinct binary features.
Finally, a finite-size driven process turns on and the system converges to the ordered phase. It is interesting to note that, in general, the binary feature reaches the convergence before the non-binary ones.


\begin{figure}[h!]
	\includegraphics[width=0.487\textwidth]{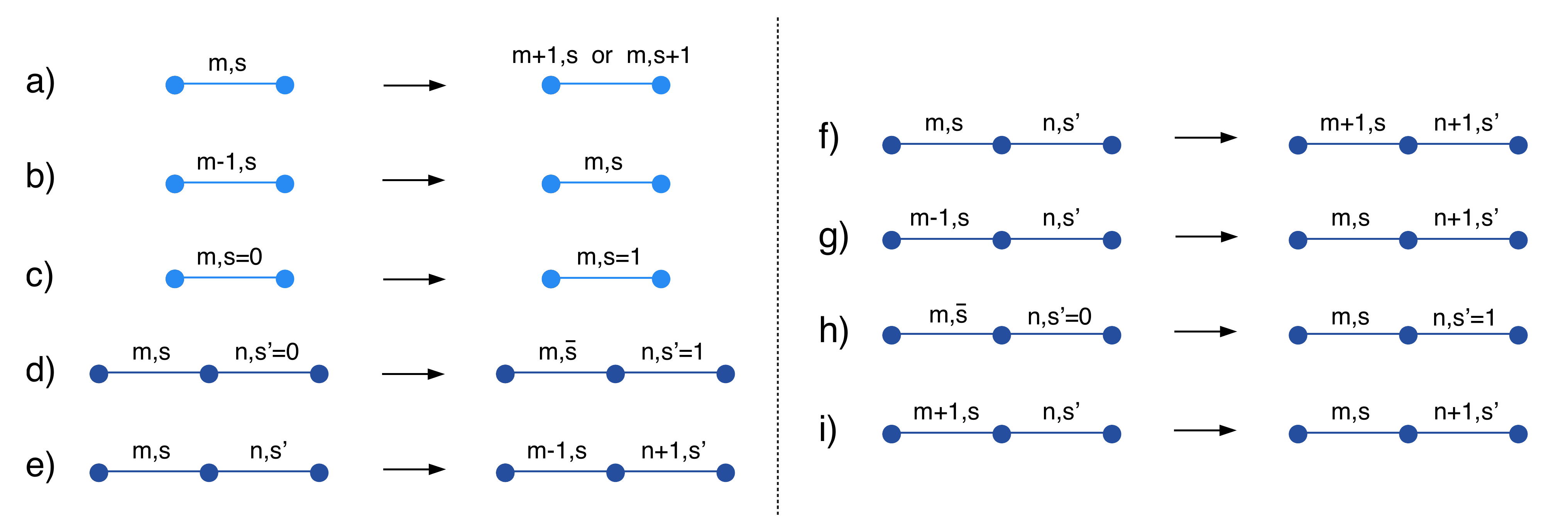} 
	\caption{Schematic description of the different processes that contribute to equation \ref{eq:MF}.
	$a)$ to $c)$ correspond to direct processes while $d)$ to $i)$ correspond to indirect ones.
	To illustrate the meaning of these diagrams lets consider two of them. 
The sub-diagram $b)$ corresponds to the direct process where a $(m-1,s)$ link evolves to a $(m,s)$ link. The probability for this process to happen is given, in a mean-field approximation, by the probability of a $(m-1,s)$ link to be sorted ($P_{m-1,s}$), times, the probability of the imitation process to happen ($m-1+s/F$), times, the probability for the imitation to occur on a non-binary feature ($(F-m+s)/(F-m+1)$), which produces the second term of equation (\ref{eq:MF}).   
The sub-diagram $e)$ corresponds to the indirect process where the state change of a link, from $(n,s')$ to $(n+1,s')$, also modifies one of the ($k-1$) neighbor links from $(m,s)$ to $(m-1,s)$. The probability for this process to occur is approximated by the probability of a $(n,s')$ link to be sorted ($P_{n,s'}$), times, the probability to have a $(m,s)$ as a neighbor link ($(k-1)\cdot P_{m,s}$), times, the probability for the imitation to happen ($(n+s')/F$), times, the probability to imitate a non-binary feature ($(F-1-n)/(F-n-s')$), times, the probability that the chosen feature was one of the $m$ shared features ($m/(F-1)$), which produce the fifth term of equation (\ref{eq:MF}). }
	\label{fig:MFterms}
\end{figure}

To gain a better understanding of the overlaps dynamics, here we extend the single link mean-field approach described in \cite{castellano2000nonequilibrium} to the case where one of the features is restricted to only two values. In order to do that, we define $P_{m,s}$ as the probability for two neighbors to coincide in $m$ features (out of $F-1$) and $s=0,1$ binary features. Then, by considering the probability of link states actualization, we can write a system of equations that describe the time evolutions of $P_{m,s}$. To compute these probabilities, we notice that when an agent $j$ imitates a feature from an agent $z$, it modifies the state of the link between $j$ and $z$ (direct process) and the state of some of the other links of agent $j$ (indirect processes). Figure \ref{fig:MFterms} displays a diagrammatic description of all the possible processes that contribute to the $P_{m,s}$  dynamics. 

By considering all the depicted diagrams we finally arrive to:


\begin{eqnarray}
\label{eq:MF}
\frac{dP_{m,s}}{dt} &=& -P_{m,s} \frac{m+s}{F} (1-\delta_{F,m}\delta_{s,1}) +\nonumber \\ & & P_{m-1,s} \frac{m-1+s}{F} \left(\frac{F-m+s}{F-m+1}\right)\nonumber + \\ & & P_{m,s-1} \frac{m}{F} \left(\frac{s}{F-m}\right)+ 
\nonumber \\
(k-1) & \times & \left[ -\sum_{n,s'} P_{m,s} P_{n,s'}\frac{n+s'}{F}\, \frac{1-s'}{F-n-s'}\right.
\nonumber \\
& & -\sum_{n,s'} P_{m,s} P_{n,s'}\frac{n+s'}{F}\, \frac{F-1-n}{F-n-s'}\,\frac{m}{F-1}
\nonumber \\
& & -\sum_{n,s'} P_{m,s} P_{n,s'}\frac{n+s'}{F}\, \frac{F-1-n}{F-n-s'}\,\frac{F-1-m}{F-1}\,\omega_{\delta(s,s')}(t)
\nonumber \\
& & +  \sum_{n,s'} P_{m-1,s} P_{n,s'}\frac{n+s'}{F}\, \frac{F-n-1}{F-n-s'}\, \frac{F-m}{F-1}\,\omega_{\delta(s,s')}(t)
\nonumber \\
& & + \sum_{n,s'} P_{m,\bar{s}} P_{n,s'}\frac{n}{F}\, \frac{1-s'}{F-n-s'}
\nonumber \\
& & + \left. \sum_{n,s'} P_{m+1,s} P_{n,s'}\frac{n+s'}{F}\, \frac{F-n-1}{F-n-s'}\, \frac{m+1}{F-1}
\right] ,
\end{eqnarray}
where $k$ is the mean connectivity of the network, $\delta(a,b)$ is the Kronecker's delta function, 
$\bar{s}$ stands for the logic negation of $s$, and $\omega_{1} (t)$ (\textit{resp}., $\omega_{0} (t)$) is the mean overlap between agents that share (not share) the binary trait. We also consider $P_{-1,s}=P_{F+1,s}=0$.





Dashed blue lines in panels \textbf{A}-\textbf{C} of Figure \ref{fig:Overlap_vs_binary_Lattice_q} show the evolution of the binary feature's overlap versus the non-binary features overlap, according to the mean-field approximation. As shown, the mean-field approximation qualitatively reproduces the behavior of all the regimes, including the first two until the finite size of the numerical simulations allows the final convergence for $q\lesssim q_c$.

\begin{figure}[h!]
	\includegraphics[width=0.3\textwidth]{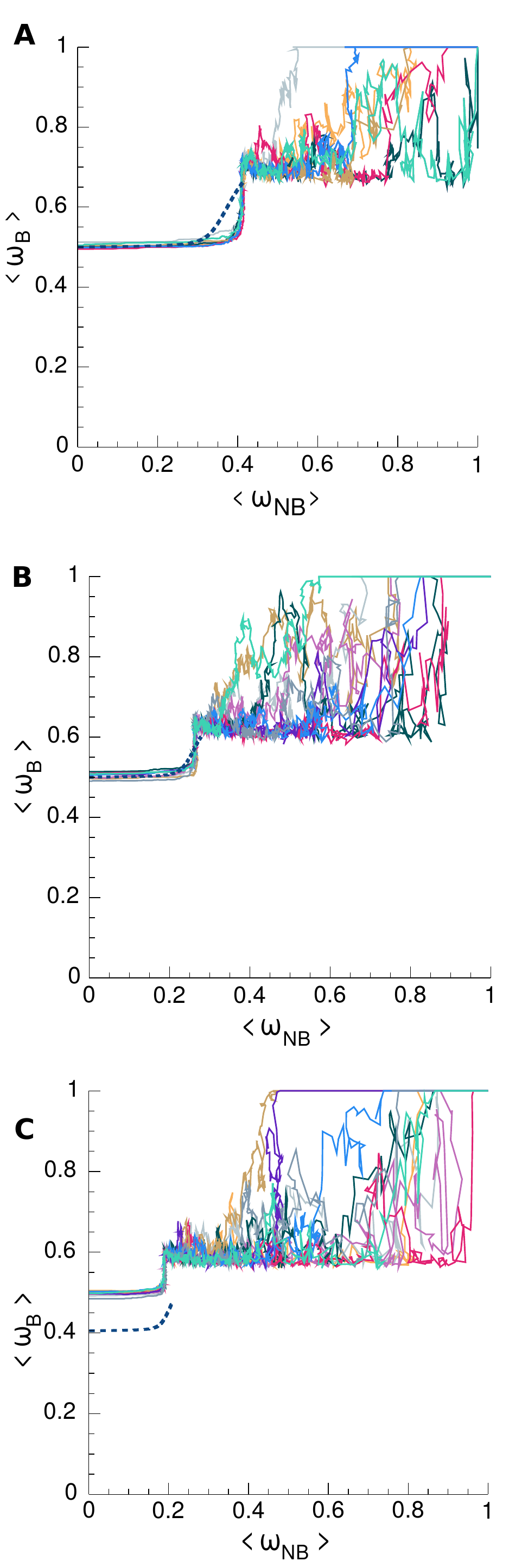}

	\caption{Solid lines show the evolution of the overlap for the binary feature versus the overlap for the non-binary features according to simulations run on a RRN ($N=1600$) with $k = 4$ (\textbf{A}), 6 (\textbf{B}), 8 (\textbf{C}); different colors correspond to 10 different characteristic runs. Dashed blue lines show  the results corresponding to the mean-field approximation. $q = 10^3$. }
	\label{fig:randomNet}
\end{figure}

To test if the mean-field description can reproduce the effect of the connectivity on the dynamical behavior, we run some simulations of our model implemented on a random regular network (RRN) with different values of the degree $k$. The thin solid lines in Figure \ref{fig:randomNet} display the numerical results for the evolution of binary and non-binary overlaps for the RRN with increasing connectivity: 
$k=4, 6, 8$ in Panels \textbf{A}, \textbf{B}, \textbf{C}, respectively. 
The thick dashed lines correspond to the mean-field approximation for the same values of $k$. As shown, the mean-field approach describes the decreasing value of the non-binary overlap reached after the first part of the dynamics, before the finite-size driven coarsening dynamics start.

As mean-field descriptions assume an infinite system size, the considered approximation can be useful also for shedding light on the absence of the transition in the thermodynamic limit, corroborating our numerical findings with some analytic arguments. If we consider a system of infinite size, as already suggested by Castellano et al. \cite{castellano2000nonequilibrium}, for $q < q_c$ the system is posed indefinitely in a coarse-grained state and, for $q > q_c$, after a characteristic time, the coarsening process stops and the
density of active links $\rho$ equals zero. Hence, for infinite systems, we can rigorously define the transition
looking at the value of $\rho$, which acts as an order parameter distinguishing among the two different
dynamical regimes: one with a perennial coarsening state ($\rho > 0$), and one with a frozen state ($\rho=0$).
Therefore, if a phase transition exists $\rho$ undergoes a discontinuity, jumping from a finite value to zero.
In our mean-field approximation $\rho$ can be computed by solving the system of equations \ref{eq:MF}, which lead to:
$\rho = \sum_{f=1}^{F-1}P_{f,s=0} +\sum_{f=0}^{F-2}P_{f,s=1}$. The corresponding stationary values of $\rho$ are independent of $q$
and always positive, without any discontinuity, implying that it is not possible to detect any phase transition in the infinite-size limit.

Finally, we study the model behavior in the regime that does not lead to a monocultural state ($q> q_c$). When the site percolation threshold of the network is below $1/2$, agents sharing a given binary trait form a giant component even for a random distribution of the traits. Therefore, initially, there will be two binary-feature clusters, one for each binary trait, allowing cultural imitation inside those components, since any pair of connected nodes belonging to one of those components has a non-zero cultural overlap. 
Those two clusters will converge separately, leading the system to a final polarized state. Conversely, when the percolation threshold of the network is above $1/2$, initially there are no giant components for the binary traits and, presumably, for large enough values of $q$ and finite system sizes, initial small clusters of agents sharing the binary feature will converge leading the system to a frozen cultural mosaic constituting a multicultural fragmented phase \cite{gracia2009residential,gracia2011selective}. Summarizing, for large values of $q$, the final state can correspond either to a fragmented phase with a large number (which scales with the system size) of different
cultures or to a polarized state, where the number of different cultures is finite, with two predominant ones.

\begin{figure}[h!]
	\includegraphics[width=0.485\textwidth]{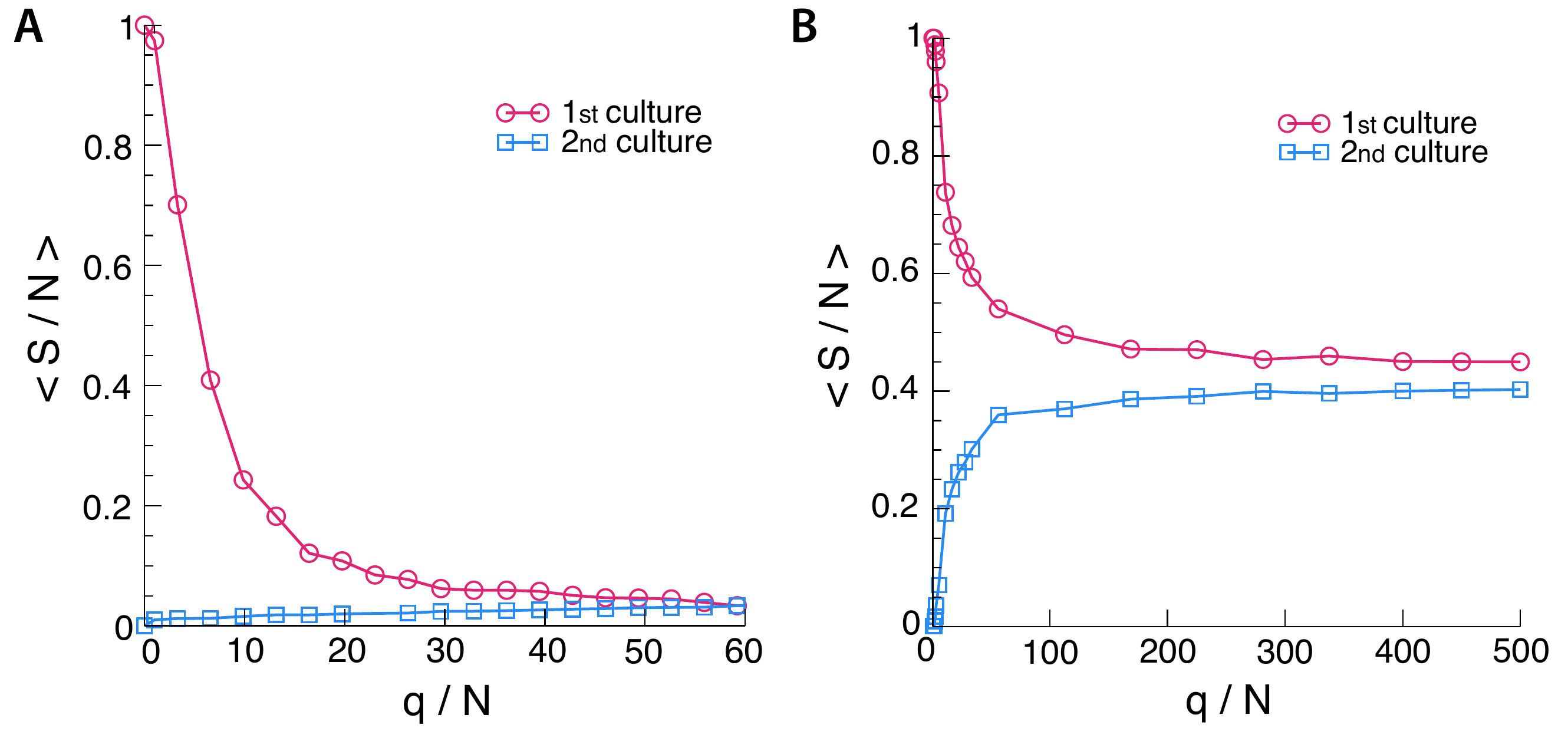}
		\caption{ Normalized size of the communities corresponding to the two largest cultures as a function of $q$ for  the model implemented on a regular square lattice (Panel \textbf{A}) and on a RRN (\textbf{B}), both with connectivity $k = 4$. Averages are taken over 100 independent simulations and $N = 1600$. }
	\label{fig:Polarize}
\end{figure}

To test this hypothesis, we have implemented the model in both a regular square lattice with connectivity $k=4$, which presents a site percolation threshold above $1/2$ 
($0.593$), and in an RRN with $k=4$, whose site percolation threshold is below $1/2$ 
($0.25$). As argued above, while the regular lattice presents a final multicultural fragmented state \cite{gracia2009residential},
when the model is implemented on a network whose percolation threshold allows the formation of giant components for the binary traits (RRN), the dynamics leads the system towards a polarized state characterized by two dominant cultures.
This behavior is described in Figure \ref{fig:Polarize}, which displays 
the normalized size ($S/N$) of the two largest cultures for both networks.
As shown in Panel \textbf{A}, for $q>q_c(N)$, in a regular lattice, the two more spread cultures cover around 10\%
of the system size. In the case of the RRN (Panel \textbf{B}), for the same $q$ values,
the two more spread communities include near 90\% of the system,
and the remaining sites correspond to a pool of different cultures organized on the border separating the two main clusters.

\section{Discussion and Conclusions}


In summary, we have presented a study of a modified version of Axelrod's model to describe the effect of cultural polarization in a population of interacting agents. This characteristic is introduced by limiting one of the cultural features to take only two possible values. The model was first analyzed on a regular 2D square lattice with the intention of clearly characterizing the nature of the transition between the ordered monocultural state and the disordered one. Our analysis shows that the introduction of a binary feature makes the well-established phase transition of the classical Axelrod's model to disappear in the thermodynamic limit.  This behavior has been characterized through a finite-size scaling analysis based on the variance of the order parameter.

These results show that the system does not display a  
genuine phase transition. 
Even if this point is not a genuine critical point, 
it has a clear physical meaning as it is the value of the $q$ 
parameter that identifies the shift from 
the order to the disorder regime.
Other models, where the transition is only observed for finite size systems, 
disappearing in the thermodynamic limit, and which present 
system size scaling, are well known in the literature \cite{toral2006finite,tessone2004neighborhood,herrero2004ising,brigatti2018exploring}.
In particular, this phenomenon has been previously observed 
in the Axelrod's model \cite{klemm2005globalization,klemm2003global,klemm2003nonequilibrium,gracia2009residential,gracia2011selective}.

In general, when we transfer statistical physics tools
to problems of social sciences, 
the population size is always  considerably smaller than the Avogadro number,
and so, not so large to justify the thermodynamic limit and its results.
In fact, we are interested in the behavior of finite-size systems, where important phenomena can appear regardless of the number of individuals \cite{toral2006finite}. For this reason, it is interesting to characterize the behavior of our system for typical finite numbers of individuals.
We have presented some results for the overlap dynamics of both binary and not binary features, displaying three different stages. In the first stage, the cultural uniformization takes place predominantly inside groups of agents with the same binary trait.
Then, only for low $q$ values, a new regime of coarsening takes place, with a first step where the cultural exchange also happens between agents, marked by distinct binary features and, finally, a finite-size driven process that leads the system to converge to the ordered phase. 
This behavior suggests that the binary feature has a predominant role in characterizing the dynamics of the system. 
This fact can be interpreted in social terms: agents find empathy on the base of some polarizing feature and then tend to align their cultures.
Consequently, the spatial structure in the initial condition of the binary feature clusters 
is essential to characterize the dynamics of the system.
Varying these initial conditions, it is possible to modify the dynamics and to
pose the system in different disordered absorbing states.
This is possible by changing the connectivity of the network of interactions.
We considered the case of a regular square lattice and random regular networks
with different average degrees. 
For the regular lattice, which presents a percolation threshold larger than $0.5$,
there is no giant component in the cluster of the binary feature and the
disordered absorbing state of the system is a fragmented one.
For the random regular networks, which present a percolation threshold smaller than $0.5$,
the clusters of the binary feature are giant components, and, therefore, the
disordered absorbing state of the system corresponds to cultural polarization. In contrast to the original Axelrod's model, always characterized by fragmentation and with a large number of different cultures scaling with the system size, our system can be partitioned in a few different cultures. 
This is an interesting result because models
with a large number of possible different absorbing states that present
polarization are uncommon \cite{brigatti2016finite,neto2020discontinuous}.
We conclude that our implementation 
can effectively describe polarization in finite populations.



\section*{Acknowledgments}
C. G. L. and Y. M. acknowledge partial support from Project No. UZ-I-2015/022/PIP, the Government of Arag\'on, Spain, FEDER Funds, through Grant No. E36-17R to FENOL, and from MINECO and FEDER funds (Grant No. FIS2017-87519-P).

\bibliography{BinaryAxelrod}{}
\bibliographystyle{ieeetr}

\end{document}